\documentclass[preprint]{aastex}

\shorttitle{Mean $JHK$ magnitudes of Cepheids}
\shortauthors{Soszy{\'n}ski et al.}

\begin{document}

\title{Mean $JHK$ Magnitudes of Fundamental-Mode Cepheids\\ from Single-Epoch Observations}

\author{I. Soszy{\'n}ski\altaffilmark{1,2}, W. Gieren\altaffilmark{1}, G. Pietrzy{\'n}ski\altaffilmark{1,2}}

\email{soszynsk@astrouw.edu.pl\\wgieren@astro-udec.cl\\pietrzyn@hubble.cfm.udec.cl}

\altaffiltext{1}{Universidad de Concepci{\'o}n, Departamento de Fisica,
  Casilla 160--C, Concepci{\'o}n, Chile}
\altaffiltext{2}{Warsaw University Observatory, Al.~Ujazdowskie~4,
  00-478~Warszawa, Poland}

\begin{abstract}
We present an empirical method for converting single-point near-infrared
$J$, $H$, and $K$ measurements of fundamental-mode Cepheids to mean
magnitudes, using complete light curves in $V$ or $I$ bands. The algorithm is
based on the template light curves in the near-infrared bandpasses. The mean
uncertainty of the method is estimated to about 0.03 mag, which is smaller
than the uncertainties obtained in other approaches to the problem in the
literature.
\end{abstract}

\section{Introduction}

In recent years great effort has been devoted to calibrating the near-infrared
(NIR) period-luminosity (PL) relationship of Cepheids. There are numerous
advantages of observing Cepheids in $J$, $H$, or $K$ bands compared to optical
observations. First, the dust extinction is an order of magnitude lower than
in the visual bandpasses. Second, the spread around the mean PL relation is
smaller. Third, metallicity effects on the PL relation are expected to
decrease in amplitude with increasing wavelength.

The largest difficulty in achieving good-quality NIR PL diagrams for Cepheids
is the necessity of obtaining well-covered light curves to compute the mean
magnitudes. However, the techniques of observations in NIR passbands are more
time-consuming, the sizes of NIR detectors are in general significantly
smaller than the optical detectors, and NIR imagers are not as widely
available on telescopes and are frequently ``bright time''
instruments. Therefore, usually only one or two random-phase observations of
Cepheids in the $J$, $H$, or $K$ bands are available. It is then necessary to
recover the mean magnitudes of the Cepheids from few-epoch data. While such
corrections to derive the mean magnitudes are usually not the dominant
remaining errors for the distance determination of galaxies from Cepheids, it
is still clearly desirable to keep these errors as low as possible.

One solution that can substitute for deriving mean magnitudes from complete
light curves is an estimation of the average brightness by template-fitting
procedures. A number of methods have been applied to reconstruct the light
curves of Cepheids in various passbands.

Historically, the first method of recovering mean $JHK$ magnitudes from
single-phase observations was described by \citet{w84}. They suggested
choosing 1 of 23 Galactic Cepheid light curves (with a period similar to the
considered variable), scaling its amplitude, and using it as a template light
curve to estimate the mean magnitude.

\citet{f88} derived the $V$ and $I$-band light curves of Cepheids in IC~1613
by scaling the amplitudes and shifting phases of the $B$-band light
curves. The adopted amplitude ratios and phase lags between photometric
bandpasses were calculated using 20 classical Galactic Cepheids. Freedman's
solution assumed that the shapes of the light curves in the various
bandpasses are the same, differing only in amplitude and phase. This
assumption works well in transformations between light curves in the optical
part of the spectrum, but is useless when transforming the optical to NIR
light curves, because the shapes of the light curves are completely different.

A series of $V$ and $I$-band template light curves were constructed by
\citet{s96}. He used more than 100 Galactic and LMC Cepheids to derive mean
Fourier coefficients of their light curves for a range of periods. The
templates were then used to recover mean magnitudes of Cepheids from a
sparsely data observed by the {\it Hubble Space Telescope} \citep{g00}.

Another solution to this problem was introduced by \citet{lst97}, who
determined the correction curves to transfer $V$ to $B$, $R$, and $I$-band
light curves of Cepheids. They used a rather small sample (six) of Galactic
Cepheids. The advantage of this method is that it does not assume fixed
shapes of the template light curves.

\citet{nge03} used statistical relations between Fourier amplitudes and
phases of $V$ and $I$-band light curves (so-called Fourier interrelations) to
reconstruct the latter. This method works properly when accurate
measurements of the Fourier coefficients to the fourth order are available;
i.e., when good quality well-covered $V$-band light curves are available.

Recently, \citet{nik04} converted random-phase $JHK$ measurements to mean
magnitudes, computing the correction functions for each bandpass. They used
more than 2000 Cepheids from the Large Magellanic Cloud (LMC) to derive the
differences between the observed magnitudes of individual variables and those
derived from fitted PL relations. The obtained residuals were then fitted by
a function depending on the $V$ phase. The weakest point of this method is
that it does not take into account the different amplitudes of individual
Cepheids. \citet{nik04} estimated the errors of their method as 0.05
mag. Similar uncertainties  for the estimation of mean $JHK$ magnitudes were
obtained by the authors of the other algorithms described above.

In this paper, we develop a technique of deriving mean NIR magnitudes of
fundamental-mode Cepheids using one random-phase measurement in $J$, $H$, or
$K$ bands and the complete light curves observed in $V$ or $I$ bands. Our
method is based on the template light curves in the $J$, $H$, and $K$ bands
obtained from 61 fundamental-mode Cepheids from the Milky Way and the LMC. The
amplitudes of our templates depend on the amplitudes in the visual passbands
($V$ or $I$ bandpasses) and on the periods of the stars. The phase lags
between visual and NIR light curves are constant in our algorithm.

Our paper is organized as follows. In Section 2 we present the development of
the $J$, $H$, and $K$ template light curves, in Section 3 we describe the
practical application of the method of deriving mean magnitudes from
single-epoch points, in Section 4 the error analysis is presented, and
Section 5 contains a summary and conclusions.

\section{Template development}

We started our analysis by selecting the most numerous possible sample of
fundamental-mode Cepheids with well-covered light curves in the NIR and visual
passbands. We used $J$, $H$, and $K$-band observations from two sources:
\citet{ls92} for Galactic Cepheids and \citet{per04} for the LMC Cepheids. The
former provide photometry in the Carter system, while the latter use $JHK_s$
magnitudes in the LCO system. Since in the ongoing Araucaria Project
\citep{g01} we calibrated our NIR observations of Cepheids in several nearby
galaxies onto the UKIRT system, we used Carpenter's (2001) transformation
equations between NIR photometric systems to transform the photometry from
both sources to this system. However, the photometric system used in our
analysis is not a crucial point, because the transformation formulae between
the systems depend only weakly on color, so they do not significantly
influence the shapes and amplitudes of the light curves.

For the selected Cepheids, we tried to find as many Johnson $V$ and Cousin
$I$-band observations in the literature as possible, to obtain complete light
curves in these bands. Ultimately, our list included 30 fundamental-mode
Cepheids from the Galaxy and 31 stars from the LMC. Tab.~1 contains the full
list of objects. The last column presents information about the sources of the
adopted data.

We collected $V$ and $I$ photometric data covering a time span of several to
more than 20 years. We used these data to correct the pulsational periods of
our sample Cepheids. In most cases the periods were measured with an accuracy
better than $5\times10^{-5} P$. We provide the improved periods in
Tab.~1. Precisely determined periods are necessary to derive correct phase
shifts between optical and NIR points.

A seventh-order Fourier series was then fitted to each $V$ and $I$ light
curve. The $J$, $H$, and $K$ light curves were fitted by a fifth-order Fourier
expansion. In this way, we obtained the basic parameters of the selected
Cepheids: magnitudes, amplitudes, and time of the maximum and minimum
brightness. In the next part of this section, we present the development of
the $JHK$ templates relative to $V$-band light curves. The analysis of the NIR
photometry relative to the $I$ bandpass was performed in the same manner.

In Fig.~1 we present the phase lags of $JHK$ light curves relative to the
$V$-band data plotted against the $\log{P}$. Note that the difference of
phases depends very weakly, if at all, on period.  In the further analysis, we
assumed that the phase lag between visual and NIR light curves does not depend
on the periods of the Cepheids.

Similarly, we tested a relationship between the amplitude ratios and the
periods. Fig.~2 presents $A_{\lambda}/A_V$ versus $\log{P}$ diagrams
($\lambda=\{J,H,K\}$). In each of the three panels, it is possible to notice
that the amplitude ratios are smaller for the short-period Cepheids and larger
for long-period Cepheids. In our algorithm, we approximated the amplitude
ratios by constant values different for shorter and longer periods of
variability. We adopted the same amplitude ratios for Galactic and LMC
Cepheids, but a different period dividing up the values. For the Galactic
Cepheids, we adopted $\log{P}=1.3$, while for the LMC we used $\log{P}=1.1$.
The difference in the amplitude ratios between Galactic and LMC Cepheids
agrees with the results of \citet{pp00}, who found statistically significant
differences in the amplitudes between Galactic, LMC, and SMC Cepheids in the
period range $1.1<\log{P}<1.4$.

Tab.~2 contains the adopted amplitude ratios for the shorter and longer period
Cepheids. Although the drop of the amplitude ratios around $\log{P} \approx
1.3$ ($\log{P} \approx 1.1$ for the LMC Cepheids) is clearly visible in
Fig.~2, the difference between shorter and longer period stars is comparable
to the scatter of the points. Therefore, the sudden adopted change of the
amplitude ratio does not have a significant influence on the final
results. One can determine and adopt another relationship of the amplitude
ratios and periods. In particular, we call attention to Cepheids with periods
longer than 100 days, for which the ratios of the NIR and $V$ amplitudes are
significantly larger than for other Cepheids.

In the next stage of our procedure, we prepared the normalized $JHK$ light
curves. For each observing point we calculated the phase from 
maximum brightness in the $V$ bandpass; i.e.,
\begin{equation}
\phi = \mathrm{mod}\left(\frac{JD^{\lambda} - JD^V_{max}}{P}\right)
\end{equation}

The magnitudes were transformed in such a way that the mean magnitude of every
light curve was 0 and the amplitude was equal to 1; e.g., for $K$-band points:
\begin{equation}
T = (K - \langle K\rangle)/A(K)
\end{equation}
\noindent
where $A(K)=K_{max}-K_{min}$ is the amplitude of variability and
$\langle K\rangle$ is a magnitude-averaged mean brightness.

All points of all light curves normalized in that manner are plotted together
in Fig.~3. The left panels show $J$, $H$, and $K$ data of the Galactic
Cepheids, while the right panels contain LMC data. Note that the
normalized light curves are very homogeneous. The scatter of the points for
the LMC Cepheids is larger than that obtained for the Galactic variables,
which is an effect of larger measurement errors for the fainter LMC objects,
but the shapes of the light curves in both environments are very similar. This
feature can be used to construct NIR template Cepheid light curves.

The last step of our analysis was an approximation of the co-added (Galactic
and LMC Cepheids) normalized light curves using a Fourier series of seventh
order:
\begin{equation}
T(\phi)=\sum_{i=1}^{7}[A_i \cos(2\pi\phi + \Phi_i)]
\end{equation}
\noindent
which were done separately for $J$, $H$ and $K$ data. The Fourier
coefficients, $A_i$ and $\Phi_i$, are presented in Tab.~3.

\section{Application of the method}

Before starting to derive the mean $\langle{J}\rangle$, $\langle{H}\rangle$,
or $\langle{K}\rangle$ magnitudes of a Cepheid, one should make sure that its
period is sufficiently well determined to accurately calculate the ephemeris
phase at the individual NIR observations. If a precise period and well-covered
light curves in $V$ or $I$ of a fundamental-mode Cepheid are available, one
can easily estimate its $\langle{J}\rangle$, $\langle{H}\rangle$, and
$\langle{K}\rangle$ using a single-epoch measurement in these filters. The
procedure is as follows.

\begin{enumerate}
\item Determine amplitudes in the visual passbands $A(V)$ or $A(I)$, defined
as the differences between the maximum and the minimum magnitude. Then using
the amplitude ratios listed in Tab.~2, estimate the amplitude in the
appropriate NIR bands.
\item Measure the epochs of maximum brightness in $V$ or $I$ bandpasses and
  calculate the appropriate phases of the NIR measurement points (equation~1).
\item Calculate the value of the template light curve $T(\phi)$ for a
  given phase using equation~3 and the Fourier coefficients listed in
  Tab.~3. The final estimation of the mean magnitudes can be obtained from the
  formulae
\begin{eqnarray}
\langle J\rangle &=& J - A(J)\times{T_J}(\phi)\nonumber\\
\langle H\rangle &=& H - A(H)\times{T_H}(\phi)\\
\langle K\rangle &=& K - A(K)\times{T_K}(\phi)\nonumber
\end{eqnarray}
\end{enumerate}

In that manner we can estimate the magnitude-averaged luminosity of the
Cepheids. However, in many cases the intensity-averaged mean magnitudes are
needed. To derive NIR intensity means expressed in magnitudes
($\langle{J}\rangle_i$, $\langle{H}\rangle_i$, and $\langle{K}\rangle_i$), we
suggest using corrections to the magnitude-averaged magnitudes. For each
fundamental-mode Cepheid in our sample, we determined mean NIR magnitudes using
both methods. Fig.~4 shows the differences between both mean magnitudes versus
the amplitude in an appropriate bandpass. Note that the difference
between magnitude-averaged and intensity-averaged mean magnitudes clearly
depends on the amplitude of variability, but stays small ($<0.015$ mag)
even for the largest amplitudes. It is possible to easily derive
intensity-averaged mean magnitudes using the
following relationships, approximated by quadratic functions:
\begin{eqnarray}
\langle{J}\rangle - \langle{J}\rangle_i & = & 0.0072 A(J) + 0.0313 (A(J))^2\nonumber\\
\langle{H}\rangle - \langle{H}\rangle_i & = & 0.0056 A(H) + 0.0329 (A(H))^2\\
\langle{K}\rangle - \langle{K}\rangle_i & = & 0.0037 A(K) + 0.0366 (A(K))^2\nonumber
\end{eqnarray}

To test our algorithm on independent data, we applied it to single-epoch $JHK$
measurements of Cepheids in the LMC from the Two Micron All Sky Survey (2MASS)
Point Source Catalog. We used the OGLE-II catalog of Cepheids in the LMC
\citep{uda99} to select fundamental-mode Cepheids, and match them with sources
from the 2MASS catalog. We found 458 counterparts closer than $2''$ from the
position of the OGLE Cepheids. We then plotted single-epoch $JHK$ measurements
in the PL diagram and removed from our sample stars that were evidently
blended in the 2MASS database. We recognized as blends objects deviating from
the mean PL relationship by more than $3 \sigma$. This seems a reasonable
assumption, given the crowding conditions in the LMC and the relatively large
$2''$ size of the pixels of the 2MASS data, which increases the number of
unresolved Cepheids compared to the optical OGLE images, which were taken at
pixel scales of $0.4''$ per pixel. Indeed, all the objects suspected of being
blends in the 2MASS data are Cepheids lying close to the respective ridge
lines in the optical $BVI$ PL relations.  We selected a total of 422
stars. The 2MASS photometry was transformed to the UKIRT system using
Carpenter's (2001) formulae.

To derive mean $JHK$ magnitudes, we utilized the well-sampled OGLE $I$-band
light curves of the Cepheids. For each object, we performed the procedure
described above. The results are presented in Fig.~5. The top panel shows
the original single-epoch measurements, while the bottom panel presents the
mean magnitudes derived from the application of our algorithm. The improvement
is clearly visible to the naked eye, especially for longer period variables
where measurement errors are smaller. For Cepheids with $P>10$ days the
scatter of the points after correction is reduced to half.

\section{Error analysis}

An estimation of the errors of the derived mean magnitudes was conducted using
the same Galactic and LMC Cepheids that served to calibrate the method. We
used each single point of each $J$, $H$, and $K$ light curve to estimate the
mean magnitudes, and we compared these values with the phase-averaged
magnitudes obtained from an integration of the complete light curves. For each
Cepheid, we calculated the $\sigma$ of the scatter of the estimated mean values
around the ``real'' mean magnitudes. The data we obtained are presented in
Tab.~4.

Obviously, the errors determined in this way are not the intrinsic errors of
our algorithm. The scatter of derived mean magnitudes also has other causes:
photometric measurement errors of the points that were used to estimate the
mean luminosity, uncertainty of the settled phases and amplitudes, and errors
of the mean magnitudes used to compare our results. The standard errors in the
individual measurements of the Galactic Cepheids were estimated to be about
0.01 mag \citep{ls92}. The typical uncertainty of the LMC data is between 0.02
and 0.05 mag \citet{per04}. Formal errors of amplitudes and mean magnitudes
are equal to 0.01--0.03 mag. The uncertainty of the phase lags between optical
and NIR light curves is a function of the period errors and time span between
observations in both bandpasses. In some cases the phase errors are as large
as 0.05. The quality of the final mean magnitudes strongly depends on the
accuracy of the ephemeris phase determination. We checked that a 0.05
inaccuracy in the phase lags increases the scatter of the derived mean
magnitudes by a factor 2.

As one can see in Tab.~4, the typical variance of the estimated mean magnitude
is about 0.03 mag for Galactic Cepheids (median values 0.035, 0.029 and 0.027
mag for $J$, $H$ and $K$ bands, respectively) and 0.05 mag for the LMC
variables (0.056, 0.046, and 0.048 mag for $J$, $H$ and $K$). The larger
scatter for the LMC Cepheids is probably an effect of larger errors of
individual measurements, compared to the Galactic Cepheid measurements of
\citet{ls92}, and the added effect of crowding on the measurements. For
several Cepheids in both galaxies, the scatter is significantly larger than
typical. We found that most of these cases are caused by atypical amplitude
ratios (e.g., for Cepheids with $P>100$ days) or phases lags between the $V$
and NIR light curves. In this are also a number of Cepheids with bumps
in their light curves.

Finally, the intrinsic error of our method of deriving mean NIR magnitudes of
fundamental-mode Cepheids can be conservatively estimated to be 0.03 mag.  The
errors can be larger when the light curves are bumpy or have atypical shapes,
but the accuracy can be better than 0.03 mag for Cepheids with typical light
curves, when the phase lags are determined with high precision.  Obviously,
the larger the number of isolated $JHK$ measurements, the better the
accuracy of the final mean magnitude, because it can be averaged from several
independent values.

\section{Summary}

The application of template light curves is the most accurate method of
estimating mean NIR magnitudes of Cepheid variables when the observed data
have poor phase coverage. We show that it is possible to reconstruct the NIR
light curves if basic parameters of the optical light curves are available.
The advantage of our method is that there is no need to determine the Fourier
coefficients for the template light curves, which for sparsely sampled light
curves cannot be measured accurately.

Obviously, the NIR templates can also be used without knowledge of the $V$ or
$I$ light curves. If several points in $J$, $H$, or $K$ bands are available
and the period of the given fundamental-mode Cepheid is known, one can fit the
proper amplitudes and phases using, for instance, the least-squares method. It
is an alternative method of converting the $J$, $H$, and $K$ magnitudes to
$\langle{J}\rangle$, $\langle{H}\rangle$, and $\langle{K}\rangle$.

\acknowledgments
{\begin{center}{\bf Acknowledgments}\end{center}
W.G. and G.P. gratefully acknowledge financial support for this work
from the Chilean Center for Astrophysics FONDAP 15010001. Support from the
Polish KBN grant 2P03D02123 and BST grant to Warsaw University Observatory
is also acknowledged.

This publication makes use of data products from the Two Micron All Sky
Survey, which is a joint project of the University of Massachusetts and the
Infrared Processing and Analysis Center, California Institute of Technology,
funded by the National Aeronautics and Space Administration and the National
Science Foundation.}

\clearpage

\begin{figure}
\includegraphics[width=16cm]{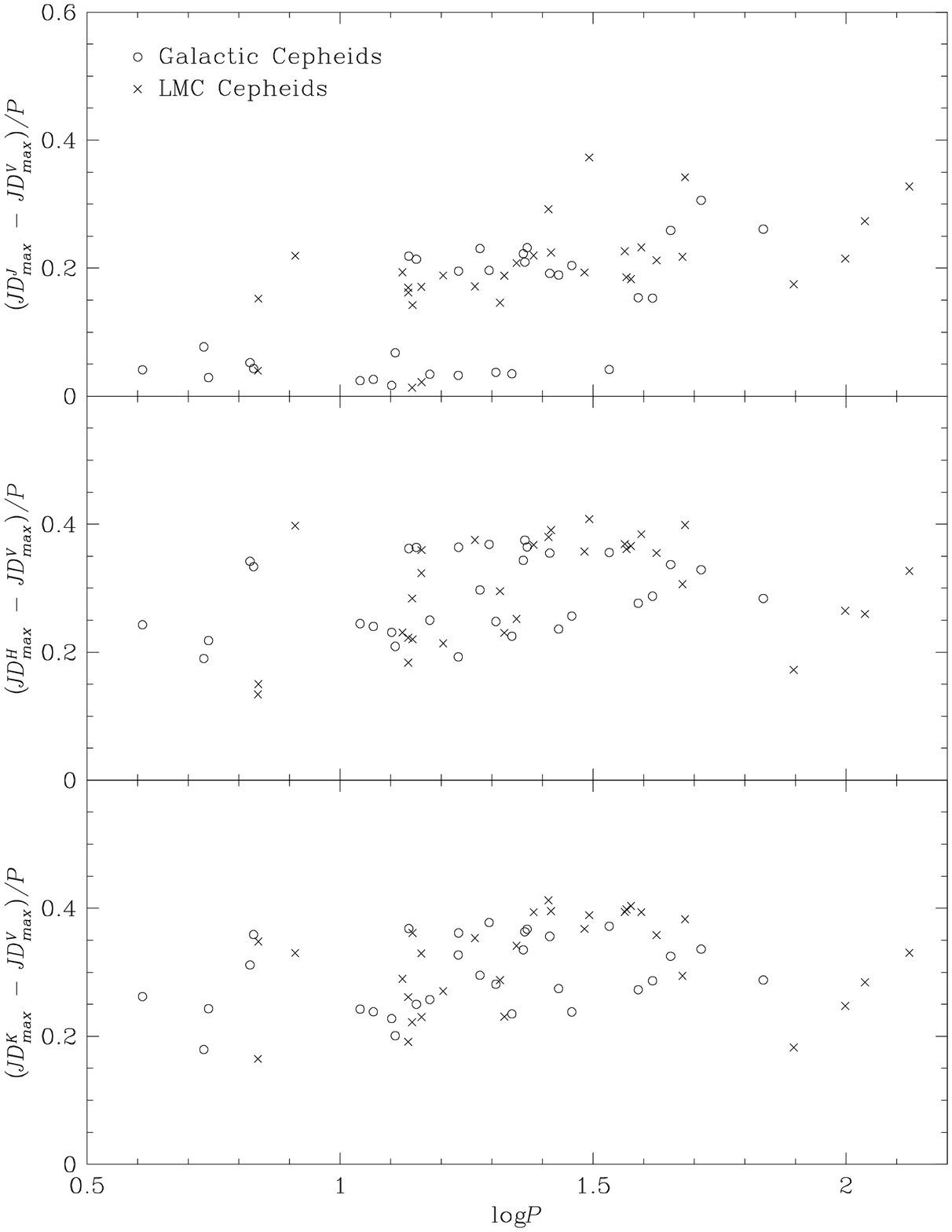}
\vspace{-1cm}
\caption{Phase lags between NIR and $V$-band light curves versus logarithm of
  the period. Circles indicate Galactic Cepheids, crosses -- LMC Cepheids.}
\end{figure}

\begin{figure}
\includegraphics[width=16cm]{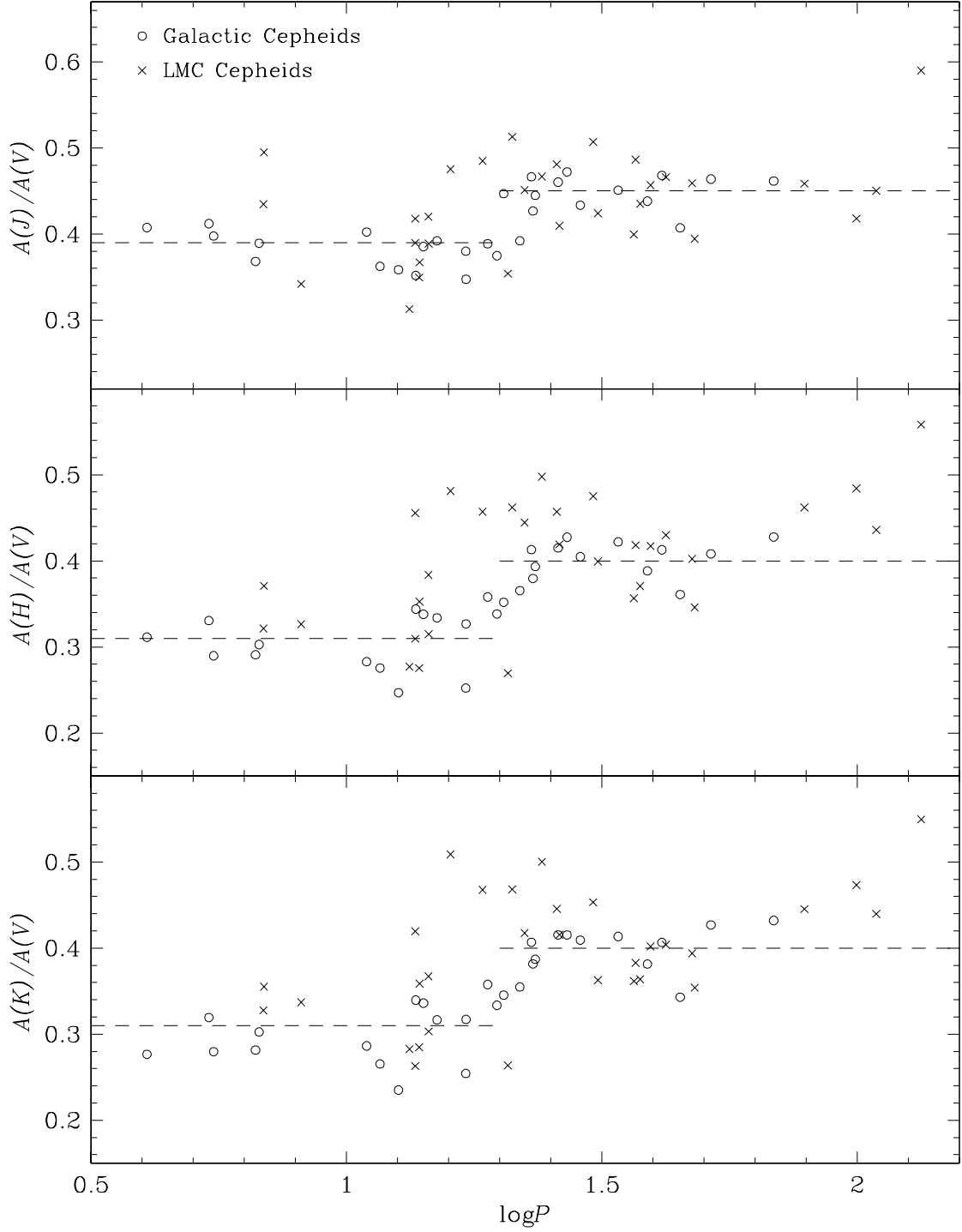}
\vspace{-1cm}
\caption{Amplitude ratios of NIR and $V$-band light curves versus the
  logarithm of the periods. Circles indicate Galactic Cepheids, crosses -- LMC
  Cepheids. Dashed lines mark adopted mean values of the amplitude ratios for
  shorter and longer-period Cepheids.}
\end{figure}

\begin{figure}
\includegraphics[width=16cm]{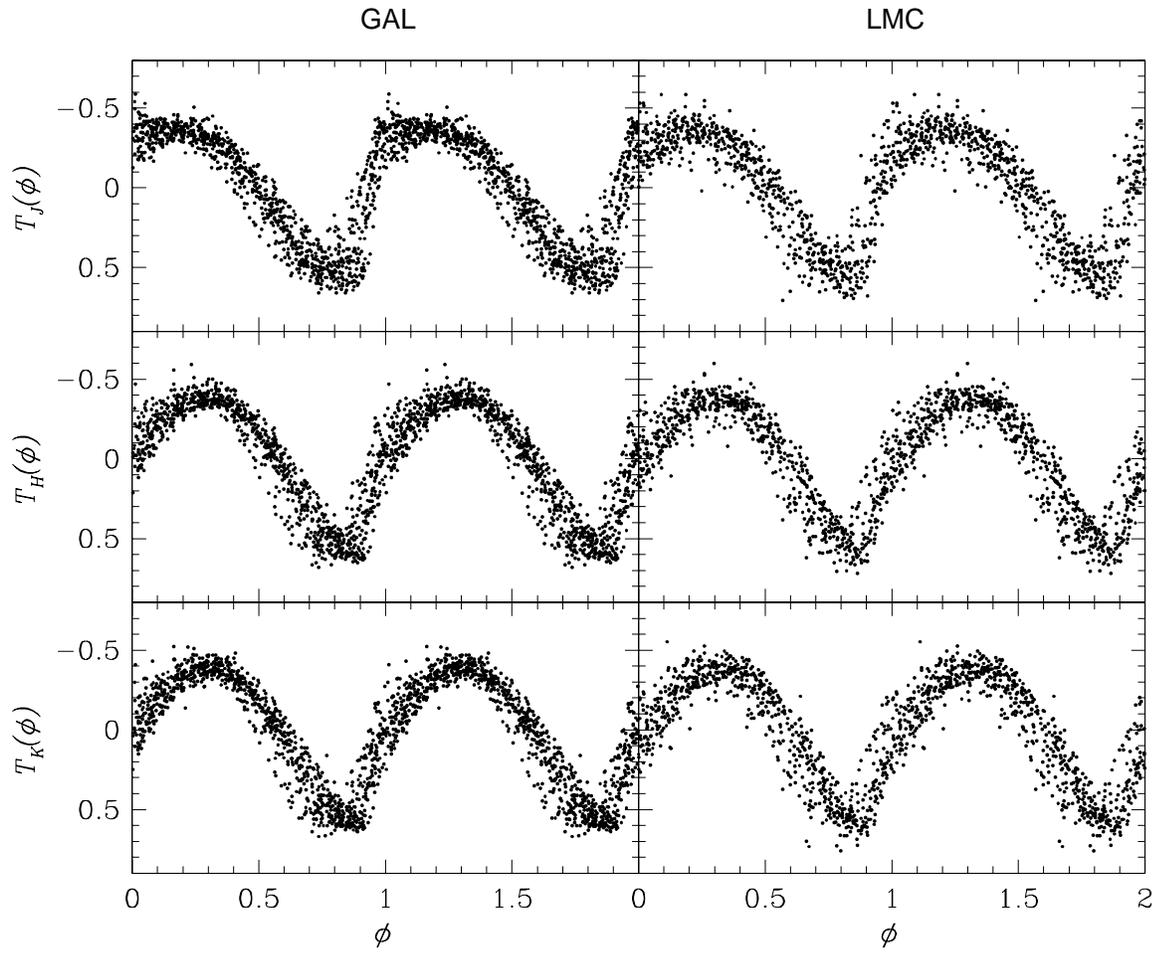}
\vspace{-8cm}
\caption{Template $J$, $H$, and $K$ light curves for fundamental-mode Cepheids
  from the Galaxy and LMC.}
\end{figure}

\begin{figure}
\includegraphics[width=16cm]{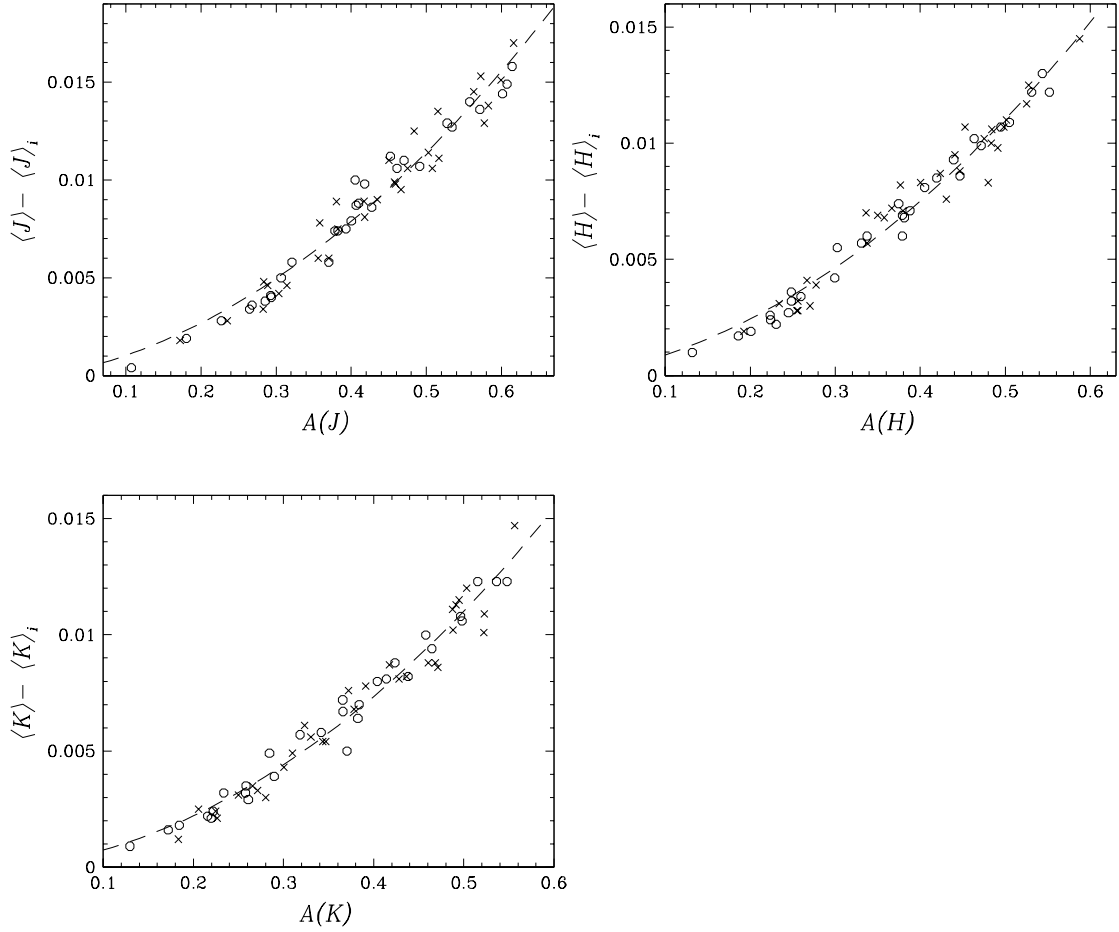}
\vspace{-8cm}
\caption{Differences between magnitude-averaged and intensity-averaged mean
  magnitudes versus amplitudes in $J$, $H$, and $K$ bands. Circles indicate
  Galactic Cepheids, crosses -- LMC Cepheids. Dashed lines are least-squares
  fits to a quadratic function.}
\end{figure}

\begin{figure}
\includegraphics[width=16cm]{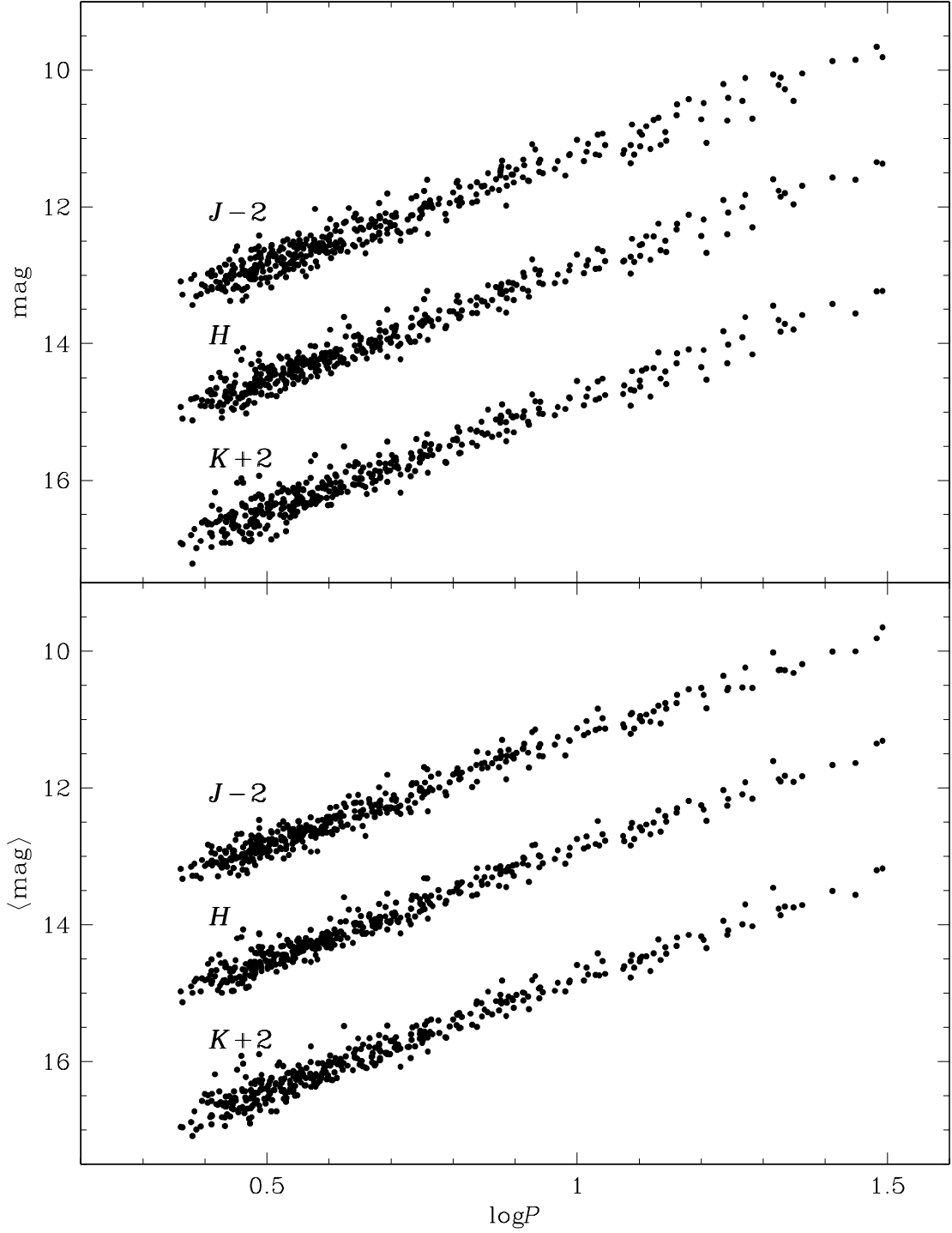}
\vspace{-1cm}
\caption{Near-infrared period--luminosity diagrams for fundamental-mode
  Cepheids in the LMC. {\it Top}: Single-epoch random-phase 2MASS photometry;
  {\it bottom}: mean magnitudes of the same stars derived by using OGLE-II
  $I$-band light curves, together with the algorithm described in this paper.}
\end{figure}

\begin{deluxetable}{crccccc}
\tablecolumns{7}
\tablewidth{0pc}
\tablecaption{List of Fundamental-Mode Cepheids Used in the Analysis.}
\tablehead{
Cepheid & \multicolumn{1}{c}{Period} & $A(V)$ & $A(J)$ & $A(H)$ & $A(K)$ & Source of \\
 & \multicolumn{1}{c}{[days]} & [mag] & [mag] & [mag] & [mag] & photometry}
\startdata
\sidehead{\bf Galactic Cepheids}
BB Sgr &  6.63712 & 0.610 & 0.225 & 0.178 & 0.172 & c,d,l\\
BF Oph &  4.06767 & 0.652 & 0.266 & 0.203 & 0.180 & c,d,l,m\\
BN Pup & 13.67253 & 1.217 & 0.428 & 0.419 & 0.413 & a,g,l,n\\
CV Mon &  5.37867 & 0.679 & 0.280 & 0.225 & 0.217 & d,l\\
GY Sge & 51.71139 & 0.593 & 0.275 & 0.242 & 0.253 & l,n\\
KN Cen & 34.02935 & 1.047 & 0.472 & 0.442 & 0.433 & g,l,m,n,o\\
KQ Sco & 28.69714 & 0.916 & 0.397 & 0.371 & 0.375 & g,m,n,o\\
LS Pup & 14.14729 & 0.994 & 0.383 & 0.336 & 0.334 & g,l,n\\
RS Pup & 41.44858 & 1.122 & 0.525 & 0.463 & 0.456 & a,d,l\\
RU Sct & 19.70259 & 1.107 & 0.415 & 0.374 & 0.369 & d,l\\
RY Sco & 20.32186 & 0.840 & 0.375 & 0.296 & 0.290 & a,d,g,m,o\\
SU Cru & 12.84922 & 0.608 & 0.107 & 0.057 & 0.051 & a,g,n\\
 S Vul & 68.62792 & 0.578 & 0.267 & 0.247 & 0.250 & l\\
SV Vul & 45.00545 & 1.049 & 0.427 & 0.379 & 0.360 & d,l,h,j\\
SW Vel & 23.42489 & 1.310 & 0.583 & 0.516 & 0.507 & g,l,m,n,o\\
SZ Aql & 17.14049 & 1.201 & 0.417 & 0.393 & 0.381 & d,l,h,n\\
 T Mon & 27.03014 & 1.010 & 0.477 & 0.432 & 0.420 & d,g,l,m,o\\
 U Car & 38.82372 & 1.186 & 0.520 & 0.461 & 0.453 & a,g,l,m,o\\
 U Nor & 12.64452 & 0.996 & 0.357 & 0.246 & 0.234 & a,g,n\\
 U Sgr &  6.74535 & 0.735 & 0.286 & 0.223 & 0.222 & c,d,l\\
UU Mus & 11.63620 & 1.074 & 0.389 & 0.296 & 0.285 & a,g\\
 V Cen &  5.49397 & 0.782 & 0.311 & 0.227 & 0.219 & c,l,m,o\\
VW Cen & 15.03728 & 0.999 & 0.392 & 0.334 & 0.317 & g,m,n,o\\
VY Car & 18.90336 & 1.042 & 0.405 & 0.373 & 0.373 & a,g,m,o\\
VZ Pup & 23.17559 & 1.297 & 0.554 & 0.492 & 0.495 & a,g,l,m,n,o\\
WZ Car & 23.01515 & 1.312 & 0.612 & 0.542 & 0.534 & a,g,l,m,n,o\\
WZ Sgr & 21.85096 & 1.127 & 0.442 & 0.412 & 0.400 & a,d,g,l,n\\
 X Pup & 25.96654 & 1.318 & 0.607 & 0.547 & 0.548 & a,d,l,m,n,o\\
XX Cen & 10.95328 & 0.907 & 0.363 & 0.255 & 0.258 & a,e,m,o\\
 Y Oph & 17.12588 & 0.494 & 0.188 & 0.125 & 0.126 & d,g,l\\
\\
\sidehead{\bf LMC Cepheids}
  HV879 &  36.82828 & 1.182 & 0.575 & 0.494 & 0.453 & b,i\\
  HV883 & 133.35058 & 1.172 & 0.692 & 0.655 & 0.644 & a,b,f,i,p\\
  HV885 &  20.70351 & 0.909 & 0.322 & 0.245 & 0.240 & k\\
  HV887 &  14.48897 & 1.088 & 0.423 & 0.343 & 0.330 & k\\
  HV889 &  25.80413 & 0.929 & 0.447 & 0.424 & 0.414 & a,k\\
  HV892 &  15.98925 & 1.014 & 0.482 & 0.488 & 0.516 & k\\
  HV893 &  21.11783 & 0.997 & 0.511 & 0.461 & 0.467 & k\\
  HV899 &  31.05049 & 1.299 & 0.551 & 0.519 & 0.471 & a,b,i,k\\
  HV900 &  47.50348 & 1.006 & 0.462 & 0.405 & 0.396 & a,b,i\\
  HV901 &  18.46964 & 0.935 & 0.453 & 0.427 & 0.437 & k\\
  HV904 &  30.39899 & 1.204 & 0.610 & 0.572 & 0.546 & k\\
  HV909 &  37.56872 & 1.166 & 0.507 & 0.432 & 0.424 & a,b,i\\
  HV911 &  13.90935 & 1.079 & 0.396 & 0.381 & 0.387 & k\\
  HV914 &   6.87847 & 0.797 & 0.346 & 0.256 & 0.261 & k\\
  HV932 &  13.28336 & 1.216 & 0.380 & 0.337 & 0.344 & k\\
  HV953 &  48.05070 & 1.050 & 0.414 & 0.363 & 0.372 & a,b,f\\
 HV1013 &  24.13171 & 0.886 & 0.414 & 0.441 & 0.443 & b\\
 HV2257 &  39.38681 & 1.217 & 0.556 & 0.508 & 0.489 & a,b,i\\
 HV2270 &  13.62556 & 0.831 & 0.347 & 0.379 & 0.349 & k\\
 HV2279 &   6.89384 & 0.897 & 0.444 & 0.333 & 0.318 & k\\
 HV2291 &  22.31693 & 1.031 & 0.465 & 0.458 & 0.430 & k\\
 HV2294 &  36.54844 & 1.351 & 0.539 & 0.482 & 0.488 & a,b,f\\
 HV2324 &  14.46634 & 0.900 & 0.378 & 0.345 & 0.330 & b,k\\
 HV2338 &  42.19749 & 1.207 & 0.563 & 0.519 & 0.488 & a,b,i\\
 HV2339 &  13.87914 & 0.974 & 0.340 & 0.268 & 0.277 & k\\
 HV2352 &  13.63052 & 0.752 & 0.293 & 0.233 & 0.198 & b,p\\
 HV2827 &  78.82556 & 0.585 & 0.268 & 0.270 & 0.261 & b,i\\
 HV2883 & 108.94612 & 1.473 & 0.663 & 0.642 & 0.648 & a,b,f,i\\
 HV5497 &  99.58612 & 0.524 & 0.219 & 0.254 & 0.248 & a,b,f,i,p\\
HV12700 &   8.15255 & 0.531 & 0.181 & 0.173 & 0.179 & b\\
HV12815 &  26.11563 & 1.155 & 0.473 & 0.484 & 0.480 & b,i\\
\enddata
\tablerefs{a -- \citet{m75}; b -- \citet{mw79}; c --
  \citet{g81}; d -- \citet{mb84}; e -- \citet{ccg85}; f -- \citet{fgm85}; g --
  \citet{cc85}; h -- \citet{bar97}; i -- \citet{mof98}; j -- \citet{kis98}; k
  -- \citet{uda99}; l -- \citet[and references therein]{bdv00}; m --
  \citet{bt01}; n -- \citet{poj02}; o -- \citet{b02}; p -- \citet{seb02}.}
\end{deluxetable}

\clearpage

\begin{deluxetable}{ccccccc}
\tablecolumns{7}
\tablewidth{0pc}
\tablecaption{Adopted Amplitude Ratios of the NIR and Visual Light Curves of Cepheids.}
\tablehead{
Period & $A(J)/A(V)$ & $A(H)/A(V)$ & $A(K)/A(V)$ & $A(J)/A(I)$ & $A(H)/A(I)$ & $A(K)/A(I)$ }
\startdata
$\log{P}<1.3$\tablenotemark{a}    & 0.39 & 0.32 & 0.31 & 0.63 & 0.50 & 0.49 \\
$\log{P}\geq1.3$\tablenotemark{a} & 0.45 & 0.40 & 0.40 & 0.70 & 0.63 & 0.62 \\
\enddata
\tablenotetext{a}{In the LMC we adopted $\log{P}=1.1$ to separate
  Cepheids with smaller and larger amplitude ratios.}
\end{deluxetable}

\clearpage

\begin{deluxetable}{ccccccc}
\tablecolumns{7}
\tablewidth{0pc}
\tablecaption{Fourier Coefficients of the Template Light Curves.}
\tablehead{
 & $T_J^V(\phi)$ & $T_H^V(\phi)$ & $T_K^V(\phi)$ & $T_J^I(\phi)$ & $T_H^I(\phi)$ & $T_K^I(\phi)$ }
\startdata
$A_1$  & 0.432  & 0.438  & 0.440  & 0.432  & 0.433  & 0.434 \\
$A_2$  & 0.109  & 0.089  & 0.082  & 0.110  & 0.094  & 0.085 \\
$A_3$  & 0.058  & 0.039  & 0.035  & 0.060  & 0.046  & 0.043 \\
$A_4$  & 0.039  & 0.019  & 0.021  & 0.038  & 0.023  & 0.020 \\
$A_5$  & 0.025  & 0.019  & 0.012  & 0.029  & 0.022  & 0.016 \\
$A_6$  & 0.016  & 0.018  & 0.013  & 0.015  & 0.021  & 0.015 \\
$A_7$  & 0.010  & 0.011  & 0.010  & 0.010  & 0.016  & 0.013 \\
\tableline
$\phi_1$  & 1.734  & 1.262  & 1.201  & 1.863  & 1.397  & 1.333 \\
$\phi_2$  & 2.747  & 2.303  & 2.174  & 2.901  & 2.543  & 2.419 \\
$\phi_3$  & 3.079  & 2.913  & 2.757  & 3.324  & 3.266  & 3.097 \\
$\phi_4$  & 3.671  & 3.394  & 3.448  & 3.934  & 3.886  & 3.794 \\
$\phi_5$  & 4.253  & 3.318  & 3.406  & 4.678  & 3.813  & 3.765 \\
$\phi_6$  & 4.460  & 3.972  & 3.658  & 5.164  & 4.380  & 4.475 \\
$\phi_7$  & 5.429  & 4.727  & 4.719  & 5.851  & 5.245  & 5.149 \\
\enddata
\end{deluxetable}

\clearpage

\begin{deluxetable}{cccc|cccc}
\tablecolumns{8}
\tablewidth{0pc}
\tablecaption{Errors of the Estimated Mean Magnitudes.}
\tablehead{
\multicolumn{4}{c|}{Galactic Cepheids} & \multicolumn{4}{c}{LMC Cepheids}\\
Cepheid & $\sigma_{\langle{J}\rangle}$ & $\sigma_{\langle{H}\rangle}$ & $\sigma_{\langle{K}\rangle}$ & Cepheid & $\sigma_{\langle{J}\rangle}$ & $\sigma_{\langle{H}\rangle}$ & $\sigma_{\langle{K}\rangle}$ }
\startdata
BB Sgr & 0.030 & 0.025 & 0.026  &    HV879 & 0.055 & 0.029 & 0.034\\
BF Oph & 0.027 & 0.021 & 0.019  &    HV883 & 0.113 & 0.097 & 0.107\\
BN Pup & 0.042 & 0.026 & 0.027  &    HV885 & 0.039 & 0.050 & 0.047\\
CV Mon & 0.025 & 0.023 & 0.020  &    HV887 & 0.049 & 0.043 & 0.042\\
GY Sge & 0.024 & 0.022 & 0.022  &    HV889 & 0.055 & 0.050 & 0.052\\
KN Cen & 0.034 & 0.025 & 0.024  &    HV892 & 0.083 & 0.073 & 0.076\\
KQ Sco & 0.017 & 0.021 & 0.029  &    HV893 & 0.050 & 0.034 & 0.037\\
LS Pup & 0.033 & 0.025 & 0.025  &    HV899 & 0.084 & 0.066 & 0.060\\
RS Pup & 0.042 & 0.029 & 0.027  &    HV900 & 0.058 & 0.042 & 0.048\\
RU Sct & 0.027 & 0.023 & 0.026  &    HV901 & 0.046 & 0.051 & 0.058\\
RY Sco & 0.018 & 0.019 & 0.020  &    HV904 & 0.079 & 0.055 & 0.051\\
SU Cru & 0.046 & 0.040 & 0.041  &    HV909 & 0.070 & 0.061 & 0.063\\
 S Vul & 0.034 & 0.030 & 0.026  &    HV911 & 0.036 & 0.028 & 0.035\\
SV Vul & 0.043 & 0.045 & 0.039  &    HV914 & 0.041 & 0.037 & 0.041\\
SW Vel & 0.089 & 0.060 & 0.056  &    HV932 & 0.044 & 0.026 & 0.029\\
SZ Aql & 0.031 & 0.016 & 0.017  &    HV953 & 0.072 & 0.056 & 0.060\\
 T Mon & 0.025 & 0.021 & 0.018  &   HV1013 & 0.051 & 0.044 & 0.050\\
 U Car & 0.048 & 0.034 & 0.034  &   HV2257 & 0.052 & 0.034 & 0.032\\
 U Nor & 0.068 & 0.057 & 0.056  &   HV2270 & 0.037 & 0.046 & 0.045\\
 U Sgr & 0.024 & 0.021 & 0.018  &   HV2279 & 0.060 & 0.046 & 0.038\\
UU Mus & 0.047 & 0.040 & 0.041  &   HV2291 & 0.037 & 0.032 & 0.027\\
 V Cen & 0.035 & 0.030 & 0.029  &   HV2294 & 0.060 & 0.046 & 0.044\\
VW Cen & 0.033 & 0.024 & 0.024  &   HV2324 & 0.034 & 0.033 & 0.033\\
VY Car & 0.053 & 0.039 & 0.039  &   HV2338 & 0.067 & 0.048 & 0.053\\
VZ Pup & 0.059 & 0.044 & 0.043  &   HV2339 & 0.068 & 0.051 & 0.051\\
WZ Car & 0.087 & 0.056 & 0.051  &   HV2352 & 0.056 & 0.037 & 0.034\\
WZ Sgr & 0.019 & 0.017 & 0.019  &   HV2827 & 0.039 & 0.041 & 0.041\\
 X Pup & 0.070 & 0.043 & 0.040  &   HV2883 & 0.094 & 0.079 & 0.083\\
XX Cen & 0.082 & 0.067 & 0.065  &   HV5497 & 0.056 & 0.059 & 0.059\\
 Y Oph & 0.036 & 0.029 & 0.027  &  HV12700 & 0.031 & 0.025 & 0.036\\
       &       &       &        &  HV12815 & 0.077 & 0.057 & 0.065\\
\enddata
\end{deluxetable}
\end{document}